\documentclass[prd,nofootinbib,showpacs,preprint,floatfix]{revtex4}
\usepackage{latexsym}
\usepackage{dcolumn}
\RequirePackage{graphicx}

\newcommand{\MET}{\slash\!\!\!\!E_T}
\newcommand{\DZero}{D$\slash\!\!\!0$}

\begin{document}

\preprint{ANL-HEP-PR-06-46}

\title{Missing heavy flavor backgrounds to Higgs boson production}

\author{Zack~Sullivan}
\thanks{Visiting scientist at Argonne National Laboratory.}
\affiliation{High Energy Physics Division,
Argonne National Laboratory,
Argonne, Illinois 60439, USA}
\author{Edmond~L.~Berger}
\affiliation{High Energy Physics Division,
Argonne National Laboratory,
Argonne, Illinois 60439, USA}

\date{June 26, 2006}

\begin{abstract}
We investigate characteristics of the signal and backgrounds for Higgs boson
decay into $WW$ at the Fermilab Tevatron and CERN Large Hadron Collider.  In
the the lepton-pair-plus-missing-energy final state, we show that the
background receives an important contribution from semileptonic decays of
heavy flavors.  Lepton isolation cuts provide too little suppression of these
heavy flavor contributions, and an additional 4 to 8 orders-of-magnitude
suppression must come from physics cuts.  We demonstrate that an increase of
the minimum transverse momentum of nonleading leptons in multilepton events
is one effective way to achieve the needed suppression, without appreciable
loss of the Higgs boson signal.  Such a cut would impact the efficiency of
searches for supersymmetry as well.  We emphasize the importance of direct
measurement of the lepton background from heavy flavor production.
\end{abstract}

\pacs{14.80.Bn, 13.85.Qk, 13.85.Rm, 13.38.-b}

\maketitle

\section{Introduction}
\label{sec:introduction}

The quest to understand the mechanism of electroweak symmetry breaking is a
central goal of high energy physics.  Experiments at the Fermilab Tevatron and
CERN Large Hadron Collider (LHC) approach this goal by searching for a new
scalar particle, the Higgs boson $H$, over a broad range of possible masses.
Electroweak precision data suggest that a standard model-like Higgs boson
should have a mass between 114 GeV and 219 GeV~\cite{LEPEWWG05}, while the
mass of a supersymmetric Higgs boson should be less than about 140 GeV.  The
measurement of Higgs boson decays into $W^+W^-$ is expected to provide the
largest significance of any final state when $135 < M_H < 219$ GeV
\cite{Han:1998sp,Carena:2000yx}.

The cleanest signature for $H\to W^+W^-$ is two isolated opposite-sign leptons
plus missing transverse energy ($\MET$) from the neutrinos in $W\to l\nu$
\cite{Han:1998sp,Carena:2000yx}.  Detailed simulations of backgrounds to this
channel from physics processes (mostly continuum $W^+W^-$ production) and fake
rates have been performed for the Tevatron~\cite{Abazov:2005un,CDFNote7152}
and the LHC~\cite{ATLTDR}.  One class of reducible backgrounds involves
processes with heavy-flavor (HF) hadrons in the final state, where at least
one lepton comes from the decay of a HF hadron (a hadron that includes either
a bottom or charm quark).  It has been assumed that these backgrounds are
removed by lepton isolation selections.  In this paper we demonstrate that
isolation does not sufficiently suppress this background.  Rather, the size of
the heavy-flavor background to lepton signatures is determined by details of
the applied physics cuts, and current analyses do not successfully remove this
background to $H\to W^+W^-$.

A rough estimate of the heavy-flavor background can be obtained from the
probability that a muon from $B$ meson decay passes isolation and basic
acceptance.  Using the PYTHIA code \cite{PYTHIA} to model production of
central ($|\eta_B|<2$) $B$ hadrons with transverse momentum $p_{TB}>10$ GeV,
and running events through the default PGS~\cite{Carena:2000yx} detector
simulation, we compute an efficiency of $8\times 10^{-3}$ for finding an
isolated muon with $p_{T\mu}>10$ GeV, and $|\eta_\mu|<2$ (the \DZero\
Collaboration independently computed an efficiency of about $5\times 10^{-3}$
in a complete detector simulation of the same process with similar
cuts~\cite{D0Note4390}).  This efficiency is dominated by the transverse
momentum spectrum of the initial $B$ hadrons, and the fragmentation function
of $B\to\mu+X$, with isolation playing a less important role.  The difference
in the rates for muons before and after \DZero-like isolation cuts shows that
isolation itself retains 10--50\% of all muons (depending on the initial $B$
momentum).  While an overall acceptance of $10^{-2}$--$10^{-3}$ is small, the
initial cross sections involving heavy flavors are many orders-of-magnitude
larger than the predicted signal.  At this level, these heavy-flavor
backgrounds would swamp or at least equal the largest backgrounds considered
so far.  It is essential to understand quantitatively whether and how
heavy-flavor backgrounds are important in the region of interest for Higgs
boson searches.

There are two classes of heavy flavor backgrounds that contribute to the
dilepton plus $\MET$ final state.  The first category includes processes that
contribute equally to like-sign lepton and opposite-sign lepton final states:
$Wb\bar b$, $Wc\bar c$, and $s$-channel single-top-quark production.  The
second class of events is more problematic in that, up to mixing effects, it
contributes only to opposite-sign lepton final states: $b\bar b$, $c\bar c$,
$Wc$, $Wb$, and $t$-channel single-top-quark production.  The $b\bar b$
process is especially worrisome, because an acceptance of $10^{-10}$ would
still leave a sizable background.

In this paper, we present a full simulation of the backgrounds for $H\to
W^+W^-\to l^+l^-{\MET}$.  We focus on two analyses: one by the \DZero\
Collaboration~\cite{Abazov:2005un} that sets a limit on the $H\to WW$ cross
section at the Tevatron, and one by the ATLAS Collaboration~\cite{ATLTDR} that
estimates their reach at the LHC.  A third analysis by the CDF
Collaboration~\cite{CDFNote7152} is also studied, but it produces neither a
Higgs boson signal nor heavy-flavor backgrounds because of extremely tight
cuts.  We limit our treatment of the CDF analysis to its potential for
suppressing some of the backgrounds at \DZero\ or ATLAS.  If a CDF analysis is
produced with sensitivity to the $H\to WW$ channel, it will encounter the same
issues as \DZero.

We begin in Sec.\ \ref{sec:d0analysis} with a description of our simulation
technique.  We then proceed with a deconstruction of each cut applied in the
\DZero\ analysis and its effect on the heavy-flavor background.  We
demonstrate that the overall $l^+l^-\MET$ background could be as much as a
factor of two larger than current estimates.  We discuss the inherent
weaknesses in a Monte Carlo estimate of the heavy-flavor background, and we
emphasize the value of direct measurements of its magnitude and kinematic
variation in the Tevatron data.  In Sec.\ \ref{sec:atlasanalysis}, we examine
this background at the LHC, following the ATLAS analysis chain in detail.  We
show that the heavy-flavor background could be overwhelming with the default
cuts.  We propose a new, more restrictive cut that would significantly reduce
the background, and we argue that the residual background can be measured
\textit{in situ}.  We conclude with a discussion of heavy-flavor lepton
backgrounds in general.

\section{Heavy flavor background at \DZero}
\label{sec:d0analysis}

The \DZero\ Collaboration searched for Higgs boson decay through $W^+W^-$ into
opposite-sign leptons~\cite{Abazov:2005un}.  We follow the complex analysis
cuts because the suppression of the heavy-flavor background (HFB) comes not
just from the isolation cuts, but from the sequence of physics cuts tuned to
extract the Higgs boson signal from the $WW$ and $Z$ backgrounds.  A side
effect of those cuts is to reduce the heavy-flavor background, but our goal is
to understand the sensitivity at each stage.

In order to make statements regarding experimental issues, we require a
detailed and believable simulation of reconstructed events.  We accomplish
this by running events through the PYTHIA 6.322~\cite{PYTHIA} showering Monte
Carlo, and feeding the output through a version of the PGS
3.2~\cite{Carena:2000yx} fast detector simulation, modified to match \DZero\
geometries, efficiencies, and detailed reconstruction procedures.  The \DZero\
physics cuts are then applied to the objects found in PGS.

The isolation criteria used by the \DZero\ Collaboration differ for muons and
electrons \cite{Abazov:2005un}.  A muon is said to be isolated if there is a
hit in the muon chamber, and the sum of the transverse momenta of tracks in a
cone $\Delta R<0.5$ around a leading track is less than $p_T^\mathrm{trk} = 4$
GeV.  The cone is defined in the plane of azimuthal angle and pseudorapidity
by $\Delta R^2 = (\Delta \phi)^2 + (\Delta \eta)^2$.  For electrons,
$E_T^{\mathrm{EM}}$ is defined to be a core cluster of transverse energy in
the electromagnetic calorimeter in a cone of size $\Delta R=0.2$.
$E_T^{\mathrm{had}}$ is the total transverse energy in a surrounding cone of
size $\Delta R=0.4$ minus $E_T^{\mathrm{EM}}$.  Isolated electrons satisfy the
requirement $E_T^{\mathrm{had}}/E_T^{\mathrm{EM}}<0.2$; and, if
$|\eta_e|<1.1$, a track must exist with $E_T^{\mathrm{EM}}/p^{\mathrm{trk}}_T
< 2$.  We apply these isolation definitions in our analysis as a modification
of PGS reconstruction.

Strong angular cuts are made in all analyses with multiple leptons.  It is
vital to maintain the correlations in our modeling.  Based on the results of
Ref.~\cite{Sullivan:2005ar}, we use MadEvent 3.0~\cite{Maltoni:2002qb} to
generate hard events, and we match the cross sections after showering to the
differential next-to-leading order (NLO) cross sections, using the procedure
described in Ref.~\cite{Sullivan:2004ie}.  For $Wjj$ and the relevant
single-top-quark events it was shown~\cite{Sullivan:2005ar} that a $K$ factor
times a leading-order (LO) distribution is sufficient to retain all angular
correlations.  The $K$ factors for $Wb\bar b$ and $Wc\bar c$ are
$K=1.5$~\cite{Sullivan:2005ar}; $K=1.5$ for $Wc$~\cite{Giele:1995kr}; and we
assume $K=1.5$ for $Wb$.  The ZTOP program provides $K=1.4$ for $s$-channel
single-top-quark production and $K=1.1$ for $t$-channel
production~\cite{Sullivan:2004ie,Sullivan:2005ar}.  We use $K=2.0$ for $b\bar
b$ production~\cite{Nason:1989zy,Olness:1997yc}, and we assume $K=2.0$ for
$c\bar c$ .  Continuum $W^+W^-$ and $H\to W^+W^-$ are evaluated from PYTHIA
routines, with $K$ factors $K=1.3$ for $W^+W^-$~\cite{MCFM} and $K=2.0$ for
$H\to WW$.  Note that the $K$ factor used for Higgs boson production is
obtained after cuts from MCFM~\cite{MCFM}, and it is larger than that used by
the \DZero\ Collaboration.

The heavy-flavor backgrounds we calculate are ultimately suppressed by at
least 5 orders-of-magnitude, and a few tricks are required to obtain a
significant sample of events for this study.  For events with a real $W$, a
selection of several hundred events is chosen that pass a loose set of cuts on
the hard matrix element: the transverse mass of the lepton and at least one
heavy quark are greater than 8 GeV, and both pseudorapidities are less than
$3.25$.  Each event is then processed by PYTHIA and PGS until two isolated
leptons are found for each cut level, or a maximum of $10^4$ trials is
reached.  This procedure provides enough events to capture the effects of
isolation and physics cuts (on phase space), and it retains the correlation
between isolation and the missing transverse energy $\MET$ cut.

Simulations of $b\bar b$ and $c\bar c$ are significantly more challenging.
Even after preselection, the cross sections change by more than 8
orders-of-magnitude.  The largest loss of events is from the $\MET$ cut, since
there is intrinsically little missing energy.  In order to obtain any events,
the procedure is modified to require only one isolated lepton.  The
probability for finding the lepton at that point in phase space is then
squared.  Cuts that depend on lepton four-vectors use the identified lepton
and any observed heavy-flavor jet in the event.  The direction of the jet is
retained, and the lepton energy fraction is estimated based on the maximum
energy that could go into pions near the lepton and still pass the isolation
energy thresholds.  The uncertainties of this method are qualitatively large
but difficult to quantify.  Given the results we obtain at the end, we
consider our simulation of $b\bar b$ a proof-of-principle that this background
must be accounted for, and not a definitive prediction of its size.

Our final results are checked against the published \DZero\ results for $H\to
W^+W^-$ and continuum $W^+W^-$~\cite{Abazov:2005un}, and they agree to within
10\%.  The good agreement with a \DZero\ isolation prediction
\cite{D0Note4390}, and a thorough study of the effect of variations of
detector parameters and reconstruction algorithms, provide some confidence
that our analysis is not very sensitive to the detector simulation.

\subsection{Numerical results}
\label{sec:d0results}

The \DZero\ analysis quotes results for the total number of opposite-sign
dileptons.  We separate the background sample into separate $ee$, $e\mu$, and
$\mu\mu$ subsamples.  In order to understand the effect of the analysis chain,
we examine each level of cuts for the $\mu^+\mu^-$ channel in
Table~\ref{tab:dzmucutlvl}.  There are two target cross sections to keep in
mind: the $H\to W^+W^-\to\mu^+\mu^-$ signal is $0.6$ fb at $\sqrt S = $~1.96
TeV if $M_H=160$ GeV; and the largest background previously estimated comes
from continuum $WW$ production.  We concentrate on backgrounds that end up
larger than the Higgs boson signal and comparable to the continuum $WW$ rate.
The single-top-quark production modes and $Wb$ end up at less than $0.25$ fb,
but they are included in the final totals below.

\begin{table*}[tbh]
\caption{Cross sections (in fb) for the $\mu^+\mu^-$ channel as a function of
cuts for the 160 GeV Higgs boson \DZero\ analysis.  Isolated muons have
$p_{T\mu}>10$ GeV, $|\eta_\mu|<2$, and satisfy \DZero\ isolation criteria.
Interval cuts summarizes the effects of three cuts described in the text that
depend on combinations of lepton momenta.
\label{tab:dzmucutlvl}}
\begin{ruledtabular}
\begin{tabular}{ldddddd}
\multicolumn{1}{c}{Cut level} & \multicolumn{1}{c}{$WW$} &
\multicolumn{1}{c}{$b\bar b$} &
\multicolumn{1}{c}{$c\bar c$} & \multicolumn{1}{c}{$Wc$} &
\multicolumn{1}{c}{$Wb\bar b$} & \multicolumn{1}{c}{$Wc\bar c$} \\ \hline
Isolated $\mu^+\mu^-$ & 62 & \multicolumn{1}{c}{$7.8\times 10^6$} & 
\multicolumn{1}{c}{$5.3\times 10^4$} & 85 & 36 & 16 \\
$p_{T_{\mu_1}} > 15$ GeV & 61 & \multicolumn{1}{c}{$5.8\times 10^6$} & 
\multicolumn{1}{c}{$3.9\times 10^4$} & 82 & 34 & 15 \\
$\MET > 20$ GeV & 49 & 208 & 5 & 51 & 19 & 7.5 \\
${\MET}_{\text{scaled}} > 15$ & 42 & 24 & \multicolumn{1}{c}{$<0.1$} & 38 &
7.7 & 4.4 \\
$H_T < 100$ GeV & 42 & 24 & \multicolumn{1}{c}{$<0.1$} & 38 & 7.7 & 4.3 \\
$\Delta\phi_{ll} < 2.0$ & 19 & 24 & \multicolumn{1}{c}{$<0.1$} & 12 & 3.3 & 1.8 \\
Interval cuts & 9.3 & 24 & \multicolumn{1}{c}{$<0.1$} & 3.1 & 2.0 & 0.9
\end{tabular}
\end{ruledtabular}
\end{table*}

The inclusive cross sections begin several orders-of-magnitude larger than
Higgs boson production.  The first level of cuts requires two opposite-sign
(OS) isolated muons with $p_{T\mu}>10$ GeV and $|\eta_\mu|<2$.  Nearly 1\% of
bottom or charm quarks has hadronized and subsequently decayed into a muon
that passes all isolation criteria.  At this point the $W+X$ cross sections
are more than 100 times that of the final Higgs boson signal, and comparable
to continuum $WW$. The $b\bar b$ and $c\bar c$ rates are 3--5
orders-of-magnitude larger.  The cut on the leading muon in the analysis is 15
GeV, but this restriction does not reduce the event rate significantly, as
shown in the Table.

The first significant physics cut is the demand for large missing energy.
This cut has little effect on $W+X$ events because there is always at least one
high-$E_T$ missing neutrino.  For $b\bar b$ or $c\bar c$ production, however,
there is a more significant suppression of the background than even the
isolation criteria.  The ${\MET}_{\text{scaled}}$ cut, where 
\begin{equation}
{\MET}_{\text{scaled}}=\frac{\MET}{\sqrt{\sum_j [\Delta E_j \sin\theta_j
\cos\Delta\phi(j,\MET)]^2}} \;,
\end{equation}
reduces sensitivity to jet mismeasurements.  It acts like a stronger $\MET$
cut that reduces $b\bar b$ and $c\bar c$ by another order of
magnitude.\footnote{A recent \DZero\ analysis \protect\cite{D0Note5063}
reduces this cut to ${\MET}_{\text{scaled}} > 7$, a reduction that
significantly weakens its power to suppress $b\bar b$ and $c\bar c$.}

The rest of the cuts are only modestly effective since they were designed for
different purposes.  A cut on the scalar sum of jet-$E_T$ ($H_T$) removes
$t\bar t$ production, but the processes that concern us rarely have additional
jets.  The $\Delta\phi_{ll}$ cut expresses the fact that spin correlations
cause the charged leptons from Higgs boson decays to align.  This cut gains a
factor of 2 reduction in the $W+X$ backgrounds, but it is not effective
against $b\bar b$.  The reason is that the $B$ meson system had to be
recoiling against missed radiation in order to pass the large $\MET$ cut.
This effect forces the leptons to be relatively close in phase space.

Finally, ``interval cuts'' refer to three different cuts that are tuned to
match features of $H\to WW$, but do not excessively impact the heavy-flavor
backgrounds here.  The cuts are
\begin{eqnarray}
20\textrm{ GeV }< M_{ll} < m_h/2 \;, \nonumber\\
m_h/2+10\textrm{ GeV}< p_{Tl_1}+p_{Tl_2}+\MET < m_h \;, \nonumber\\
m_h/2 < M_T^{ll}< m_h-10\textrm{ GeV} \;.
\end{eqnarray}
Variable $M_{ll}$ is the invariant mass of the dilepton pair, and the 
transverse mass is 
\begin{equation}
M_T^{ll}=\sqrt{2{\MET} p_T^{ll}[1-\cos(\phi_{\MET} - \phi_{ll})]} \;.  
\end{equation}

Since at least one of the muons comes from a heavy flavor decay, it tends to
be relatively soft.  This muon is soft \textit{only} because the initial
heavy-flavor hadron tends to be soft, and \textit{not} because it is taking a
small fraction of the hadron's momentum.  In most of these events, the muon
has acquired most of the hadron momentum, leaving little surrounding energy in
the event on which to cut.

The cut on the sum of the transverse momenta, the second line in the list
above, is almost automatically satisfied by phase space considerations.  Its
largest effect is on the $Wc$ process, where the charm spectrum is fairly soft
and so many events fall below the minimum.  The transverse mass cut is
essentially always satisfied over the large interval allowed.  We examine this
distribution in detail later.

The final cross sections for a 160 GeV Higgs boson, the $W^+W^-$ continuum,
and the heavy-flavor backgrounds are listed in Table~\ref{tab:dzemu} for each
final state.  Statistical uncertainties are included for the $W+X$
backgrounds, while the $b\bar b$ and $c\bar c$ uncertainties are large and
unknown.  We list both like-sign and opposite-sign lepton channels to
demonstrate that only some of the backgrounds are symmetric.  If the
backgrounds were entirely driven by parton-level physics, then there would be
no like-sign contribution from $Wc$, since only $W^+\bar c$ or $W^-c$ are
produced, and mixing is small.  However, there is some probability of tagging
a jet (especially a wide-angle pion) as an electron, and hence there is an
underlying contribution to both LS and OS samples in the $ee$ and $e\mu$
samples, but not the $\mu\mu$ sample.  This contamination is small but within
the same order of magnitude of an isolated lepton from the charm decay.

\begin{table*}[tbh]
\caption{Detailed cross sections (in fb) for like-sign (LS) and
opposite-sign(OS) leptons after all cuts for the \DZero\ analysis tuned for a
160 GeV Higgs boson.  Cross sections less than $0.25$ fb are summed under all
else.  $c\bar c$ contributes $1.4$~fb to OS $ee$. Statistical uncertainties are
shown where available.
\label{tab:dzemu}}
\begin{ruledtabular}
\begin{tabular}{ccccccc}
& \multicolumn{2}{c}{$ee$} & \multicolumn{2}{c}{$e\mu$} &
\multicolumn{2}{c}{$\mu\mu$} \\
& \text{LS} & \text{OS} & \text{LS} & \text{OS} & \text{LS} &
\text{OS} \\ \hline
$H\to WW$ & \text{---} & $0.73\pm 0.04$ & \text{---} & $1.26\pm 0.05$ & \text{---} & $0.60\pm 0.03$ \\ \hline
$WW$ & \text{---} & $12\pm 1$ & \text{---} & $20\pm 1$ & \text{---} & $9.3\pm 0.9$ \\
$b\bar b(j)$ & \text{---} & 2.1 & \text{---} & 5.6 & \text{---} & 24 \\
$Wc$ & $0.8\pm 0.4$ & $2.3\pm 1.1$ & $1.1\pm 0.4$ & $3.7\pm 1.8$ & \text{---} & $3.1\pm 2.2$ \\
$Wb\bar b$ & $0.4\pm 0.2$ & $0.4\pm 0.1$ & $2.1\pm 1.6$ & $1.3\pm 0.4$ & $2.5\pm 1.6$ & $2.0\pm 1.1$ \\
$Wc\bar c$ & $1.4\pm 0.5$ & $1.1\pm 0.4$ & $1.0\pm 0.2$ & $1.6\pm 0.3$ & $1.0\pm 0.4$ & $0.9\pm 0.2$ \\
all else & $0.1$ & $1.6$ & $0.3$ & $0.3$ & $0.04$ & $0.1$
\end{tabular}
\end{ruledtabular}
\end{table*}

In Table \ref{tab:dzemu} the row containing $b\bar b$ production is modeled by
two hard processes: $b\bar b$ and $b\bar bj$, where $j$ stands for an extra
jet with $E_{Tj}>20$ GeV.  The isolated muon sample comes almost entirely from
$b\bar b$, while the isolated electron sample comes almost entirely from
$b\bar bj$.  More electrons appear in the $b\bar bj$ sample than the $b\bar b$
sample because the $B$ hadrons recoiling against the jet are harder than
estimated by $b\bar b$ plus showering.  Hence, more electrons pass the minimum
$E_T$ cut in the isolated electron definition.  Sometimes the additional jet
is missed and helps the events pass the missing energy cut.  While the
additional radiation does add some sensitivity to the $H_T$ and mass-window
cuts, it is not enough to observe in the Table.  Because of the way the
calculation is performed, we do not include the effect of finding one lepton
from the $b$ decay, and one fake $e$ from additional jets.

The results in Tables \ref{tab:dzmucutlvl} and \ref{tab:dzemu} imply that the
$l^+l^-\MET$ final state for a 160 GeV Higgs boson has an additional
background of 16 events from heavy-flavor decays ($W+X$ plus $b\bar b$) in 330
pb$^{-1}$ of data, compared to the combined background of $20$ events
estimated by \DZero\ (dominated by 12 events from continuum
$WW$)~\cite{Abazov:2005un}.  The uncertainty of the contribution from $b\bar
b$ is large, but even if the $b\bar b$ contribution to the background is
overestimated by a factor of 10, the remaining heavy-flavor backgrounds (from
$W+X$) are fully half as large as the continuum $WW$ rate.

In Table \ref{tab:dzvmh}, we demonstrate that the heavy-flavor background is
significant in each final state ($ee$, $e\mu$, $\mu\mu$) across the entire
range of Higgs boson masses studied by \DZero.  At this point we might be
concerned that there is no apparent excess in the data.  However, there is
significant uncertainty in the overall normalization of the background.  Even
a doubling of the background is consistent within 1--2$\sigma$ when the
systematic uncertainties are included.  Nevertheless, given the uncertain
nature of modeling tails of distributions, it is clear that the relative
importance of the backgrounds will only be disentangled by a direct
measurement of the heavy-flavor component.

\begin{table*}[tbh]
\caption{Cross sections (in fb) at the Tevatron for the $H\to WW$ signal and
heavy-flavor backgrounds (HFB) for each pair of opposite-sign leptons as a
function of Higgs boson mass after \DZero-like analysis cuts.  Continuum $WW$
is also shown for comparison.
\label{tab:dzvmh}}
\begin{ruledtabular}
\begin{tabular}{cddddddddd}
& \multicolumn{3}{c}{$ee$} & \multicolumn{3}{c}{$e\mu$} &
\multicolumn{3}{c}{$\mu\mu$} \\
$m_h$ (GeV) & \multicolumn{1}{c}{$H\to WW$} & \multicolumn{1}{c}{HFB} & \multicolumn{1}{c}{$WW$} & \multicolumn{1}{c}{$H\to WW$} & \multicolumn{1}{c}{HFB} & \multicolumn{1}{c}{$WW$} & \multicolumn{1}{c}{$H\to WW$} & \multicolumn{1}{c}{HFB} & \multicolumn{1}{c}{$WW$} \\ \hline
120 & 0.15 & 13 & 7.3 & 0.22 & 23 & 11 & 0.09 & 34 & 5.4 \\
140 & 0.47 & 12 & 10 & 0.90 & 20 & 16 & 0.41 & 32 & 8.4 \\
160 & 0.73 & 7.4 & 12 & 1.26 & 12 & 20 & 0.60 & 30 & 9.3 \\
180 & 0.53 & 5.9 & 11 & 0.88 & 9.8 & 18 & 0.45 & 26 & 9.3 \\
200 & 0.23 & 4.8 & 8.9 & 0.41 & 7.4 & 16 & 0.19 & 25 & 8.2
\end{tabular}
\end{ruledtabular}
\end{table*}
% The following use K(H) = 1.0
%120 & 0.07 & 13 & 7.3 & 0.11 & 23 & 11 & 0.05 & 34 & 5.4 \\
%140 & 0.23 & 12 & 10 & 0.45 & 20 & 16 & 0.21 & 32 & 8.4 \\
%160 & 0.37 & 7.4 & 12 & 0.63 & 12 & 20 & 0.30 & 30 & 9.3 \\
%180 & 0.27 & 5.9 & 11 & 0.44 & 9.8 & 18 & 0.23 & 26 & 9.3 \\
%200 & 0.12 & 4.8 & 8.9 & 0.21 & 7.4 & 16 & 0.09 & 25 & 8.2

\subsection{Measuring the background}
\label{sec:measureb}

The results of our analysis demonstrate two points: despite small
efficiencies, heavy-flavor decays into leptons are a potentially serious
background; and the efficiencies are so small, it is difficult to believe any
absolute predictions based on Monte Carlo techniques.  Therefore, the
background must either be measured \textit{in situ}, or the cuts must be made
more restrictive in order to avoid the problem.  The latter case will be
examined in Sec.\ \ref{sec:d0newcut}, but here we examine whether it is
possible to measure the background.

Let us return to the two classes of backgrounds from heavy flavor decays.  The
first class involves processes that have a roughly equal probability of
producing like-sign and opposite-sign leptons.  These include $Wb\bar b$,
$Wc\bar c$, and $s$-channel single-top-quark production.  The simplest choice
to measure the backgrounds in this case is to measure the like-sign leptons
and use this result as a measure of the background to opposite-sign leptons.
The experimental challenge is to accurately predict small variations in
efficiencies for the two final states.  Nevertheless, this procedure should
give a reasonable estimate of the background.

The second class of backgrounds is challenging to measure and, as is evident
in Table \ref{tab:dzemu}, a larger percentage of the total background.  There
are several processes that contribute only to the opposite-sign final state
(up to heavy-flavor meson mixing effects and some wrong-sign charm decays).
For a first estimate of the effect of isolation cuts, one could look at a muon
triggered sample for a tagged $b$($c$).  If a cut is made on missing
transverse energy of $\MET>10$--15 GeV, it should be possible to observe two
peaks in the muon $p_T$ spectrum of the $\mu+b$ sample. $B\bar B$ production
will peak at the muon $p_T$ threshold and exhibit a long tail to larger $p_T$,
while $Wb+X$ will peak near 40 GeV.  Once isolation criteria are imposed on
this sample, an upper bound on the effect of isolation can be
obtained.\footnote{The charm and dijet mistag background should be a small
fraction of the total events.}

While the muon sample should be fairly clean, the electron plus $b$ sample may
be more sensitive to fakes at lower energies.  The $e+b$ sample will
contain an enhanced fraction of $b\bar b$($c\bar c$), but there will be
additional backgrounds from $Z\to\tau^+\tau^-$, top-quark production, dijets,
and $Wj$ events that must be understood.

It is more complicated to estimate the $b\bar b$ or $c\bar c$ processes
because missing energy from escaping neutrinos can help these processes pass
the $\MET$ cut.  An observation of the acceptance as a function of $\MET$ may
help to reduce some background and partially separate $b\bar b$ from
$Wb/Wc/Wb\bar b/Wc\bar c$.  Unfortunately, these backgrounds are always at the
tails of the distributions, and this sort of study is more useful for
understanding general physics properties in a detector than for measuring the
background to Higgs boson production.

\begin{figure}[tbh]
\centering
\includegraphics[width=3.25in]{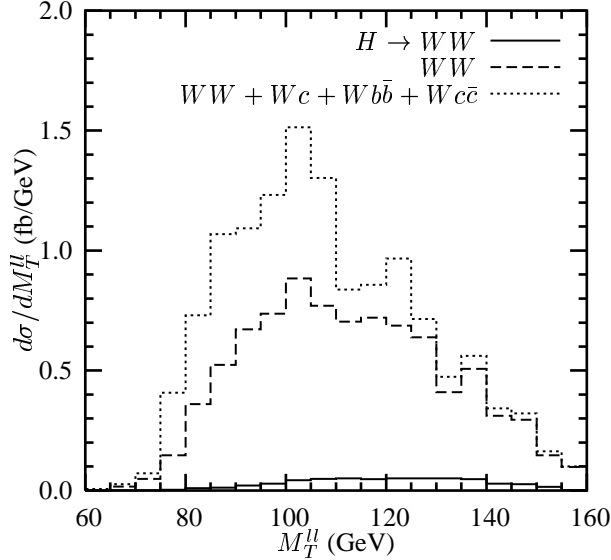}
\caption{Opposite-sign dilepton transverse mass distribution for a 160 GeV
Higgs boson, the continuum $WW$ background, and the sum of all $W+X$
backgrounds with \DZero~analysis cuts.  The $b\bar b$ and $c\bar c$
contributions are not included, but they are expected to peak around 80--100
GeV with unknown tails.
\label{fig:dzmtll}}
\end{figure}

One handle on the background to Higgs boson decay would be to loosen the cuts
as little as possible until a more pure sample of background is obtained, and
assume that the procedure can be reversed.  In Fig.~\ref{fig:dzmtll}, we see
the transverse mass distributions for $H\to WW$, continuum $WW$, and continuum
$WW$ plus the heavy-flavor backgrounds that involve a real $W$.  Our
statistical sample of $b\bar b$ events is too small to predict the shape here
(but see the ATLAS analysis in Sec.\ \ref{sec:atlasanalysis}).  The most
significant characteristics of the HF backgrounds are that they tend to peak
at a slightly lower $M_T^{ll}$ and have narrower distributions than continuum
$WW$.  If enough data are collected, and the shape of continuum $WW$ from
Monte Carlo is correct\footnote{The residual Drell-Yan and $W+$fake
backgrounds must also be included.}, one can try to make an \textit{in situ}
measurement of the heavy-flavor background.  For example, one might look at
events above and below 110--120 GeV, and set a limit on the size of the
combined heavy-flavor background.  A reasonable measurement might be possible
with a few inverse femtobarns of data.

\subsection{Cutting away the background}
\label{sec:d0newcut}

Given the challenge of measuring the heavy-flavor background, the question
arises whether it can simply be cut away.  One source of inspiration could be
the CDF analysis \cite{CDFNote7152}, but that method also
suppresses the signal, so we settle for varying the cuts at each stage
of the analysis.  The first place to look would be to reexamine the isolation
criteria, but it is difficult to achieve order-of-magnitude
suppressions by varying isolation.

Our modest simulations suggest that the HF leptons passing isolation cuts fall
into two categories.  In the first category, the hadron remnant is outside the
\DZero\ isolation cone $\Delta R>0.5 (0.4)$ for muons (electrons) and so is
not counted.  An increase of the isolation cone size to $0.7$ achieves only a
factor of 2 reduction of the backgrounds, but it begins to eat into the good
lepton sample we wish to retain.  The second category is more problematic.  In
this case, the hadron remnant is too soft to fail the 4 GeV isolation energy
threshold for muons, or the $E_{\mathrm{had}}/E_{\mathrm{EM}}<0.2$ cut for
electrons.  These events are more difficult to reject, because the thresholds
cannot be lowered by much before sensitivity to the underlying event or
minimum ionizing radiation (for muons) increases.  More importantly, these
events come from the portion of phase space where the lepton takes most of the
visible hadron momentum.  Even the impact parameter is small.

\begin{figure}[tbh]
\centering
\includegraphics[width=3.25in]{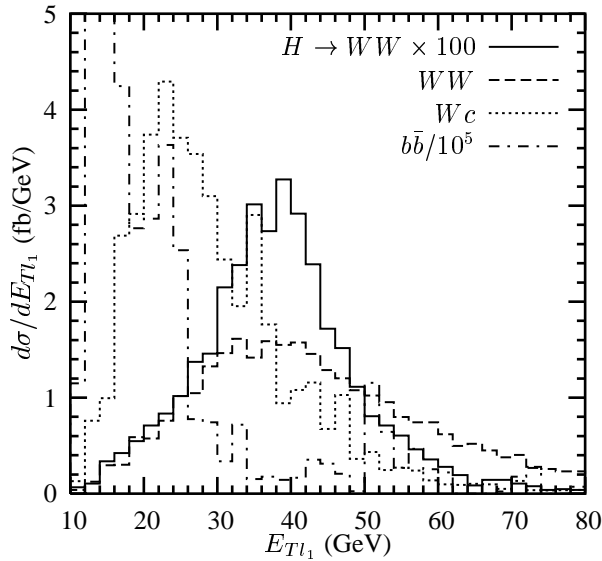}
\caption{Transverse energy distribution of the leading muon after isolation, 
with \DZero\ analysis cuts.
\label{fig:dzetlo}}
\end{figure}

\begin{figure}[tbh]
\centering
\includegraphics[width=3.25in]{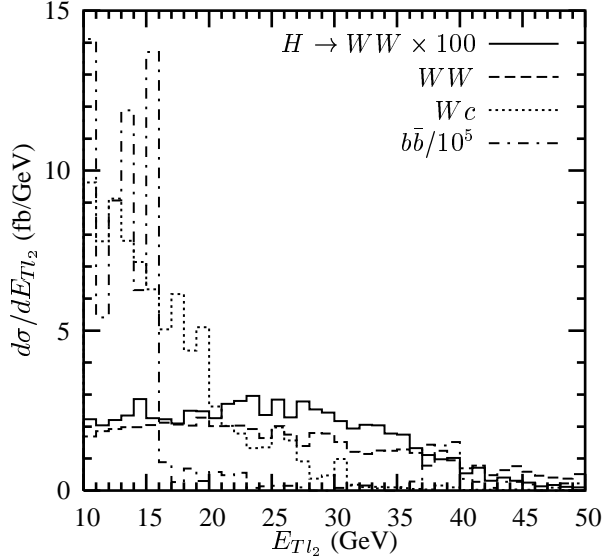}
\caption{Transverse energy distribution of the next-to-leading muon after
isolation, with \DZero\ analysis cuts.
\label{fig:dzetlt}}
\end{figure}

A better method of cutting the background is to look at the transverse energy
$E_{Tl}$ spectra of the leptons.  In Figs.\ \ref{fig:dzetlo} and
\ref{fig:dzetlt}, we see the $E_{Tl}$ of reconstructed leptons for $H\to WW$,
continuum $WW$, $Wc$, and $b\bar b$.  An increase of the cut on the leading
lepton from 15 GeV to 20 GeV is useful, but not as beneficial as it might
seem, because the $Wc$ leptons peak above 20 GeV, and even the leading lepton
from $b\bar b$ is not falling very fast.  However, the spectrum of the
next-to-leading lepton in Fig.\ \ref{fig:dzetlt} is falling exponentially,
whereas the Higgs boson and continuum $WW$ spectra are almost flat up to 40
GeV.  An increase of the cut on the next-to-leading lepton close to that of
the leading lepton (e.g., make both 20 GeV) virtually removes the $Wc$ and $b
\bar{b}$ backgrounds, with a modest reduction of the signal.  Of course this
increase provides no help against the original continuum $WW$ background.

While it appears that there is a way to effectively remove the background from
heavy-flavor leptons for the $H\to WW$ search at the Tevatron, it is not
obvious that it is desirable.  Given the poor signal to background ratio of
about $1/30$ before the addition of the HF backgrounds, we would recommend
that more effort be placed on understanding how to extract the HF background
from the data.  That understanding could then be fed into other multilepton
analyses and lay the foundation for studies at the LHC.

\section{Heavy flavor background at ATLAS}
\label{sec:atlasanalysis}

The Higgs boson is expected to be discovered with a large significance at the
LHC~\cite{ATLTDR}.  In this Section, we examine the signal and backgrounds for
$H\to WW \to l^+l^-\MET$ in the ATLAS detector, emphasizing the
contribution from leptons that arise from semileptonic decays of heavy
flavors.

The simulation procedure is the same as that used for the \DZero\ analysis in
Sec.\ \ref{sec:d0analysis}.  The detector simulation is a heavily modified
version of PGS~\cite{Carena:2000yx} that reproduces the results of the
relevant full ATLAS detector simulations to within 10\%~\cite{ATLLEPS}.  This
code has a more accurate treatment of geometric effects and efficiencies than
ATLFAST.  One surprise is that effective $K$ factors {\em after cuts} are
uniformly smaller at the LHC than at the Tevatron.  Using MCFM 4.1~\cite{MCFM}
and ZTOP 1.0~\cite{Sullivan:2004ie,Sullivan:2005ar}, we find the $K$ factor
for $H\to WW$ is only $1.25$ (compared to $2$ at the Tevatron); $K = 1.5$
for $s$-channel single-top-quark production; and all other $K$ factors are
approximately $1.0$.  The net effect is that only the Higgs boson signal is
(slightly) enhanced at NLO.

There are two caveats that must be noted.  First, the lepton identification
and isolation cuts will likely change when data are accumulated and detector
response is measured.  The large inherent uncertainties in the simulation
procedure will not be improved until the detectors are operating.
Nevertheless, the results are dominated by detector-independent physics and
should be fairly accurate relative to other physics processes.

The second caveat is that we are able to calculate only a \textit{lower} bound
on the $b\bar b$ background, and we do not include the $c\bar c$ background.
The reason is that the ATLAS study imposes extremely tight cuts on the phase
space of the leptons that are passed only if strong preselections are made on
the events.  Rough estimates indicate that the real background will be at
least a factor of 2--3 larger than presented here.  However, we demonstrate
that the shape is more of a limiting factor, and we can still draw conclusions
regarding the heavy flavor backgrounds.

In order to compare to the ATLAS study, we follow the cuts described in the
ATLAS Technical Design Report (TDR)~\cite{ATLTDR}.  These cuts are similar to
those used by \DZero, but they have tighter restrictions on angular variables
and reconstructed masses.  The definitions of isolation are involved,
particularly for electrons, and are spelled-out in
Refs.~\cite{ATLTDR,ATLLEPS}.  In Table \ref{tab:atlcutline} we show the effect
of each level of cuts on the opposite-sign dilepton events.

\begin{table*}[tbh]
\caption{Cross sections (in fb) for opposite-sign leptons as a function of
cuts for the 160 GeV Higgs boson ATLAS analysis.  $b\bar bj^{\star}$ production is a
lower limit based on limited phase space, and $c\bar c$ production is not
calculated.  A dash indicates statistics were too small to estimate.
\label{tab:atlcutline}}
\begin{ruledtabular}
\begin{tabular}{lddddddd}
\multicolumn{1}{c}{Cut level} & \multicolumn{1}{c}{$H\to WW$} &
\multicolumn{1}{c}{$WW$} &
\multicolumn{1}{c}{$b\bar bj^\star$} &
\multicolumn{1}{c}{$Wc$} &
\multicolumn{1}{c}{single-top} &
\multicolumn{1}{c}{$Wb\bar b$} & \multicolumn{1}{c}{$Wc\bar c$} \\ \hline
Isolated $l^+l^-$ & 336 & 1270 & \multicolumn{1}{r}{$>35700$} & 12200 & 3010 & 1500 & 1110 \\
$E_{Tl_1}>20$ GeV & 324 & 1210 & \multicolumn{1}{r}{$>5650$} & 11300 & 2550 & 1270 & 963 \\
$\MET > 40$ GeV & 244 & 661 & \multicolumn{1}{r}{$>3280$} & 2710 & 726 & 364 & 468 \\
$M_{ll} < 80$ GeV & 240 & 376 & \multicolumn{1}{r}{$>3270$} & 2450 & 692 & 320 & 461 \\
$\Delta\phi < 1.0$ & 136 & 124 & \multicolumn{1}{r}{$>1670$} & 609 & 115 & 94 & 131 \\
$|\theta_{ll}|<0.9$ & 81 & 83 & \multicolumn{1}{r}{$>1290$} & 393 & 68 & 49 & 115 \\
$|\eta_{l_1}-\eta_{l_2}| < 1.5$ & 76 & 71 & \multicolumn{1}{r}{$>678$} & 320 & 48 & 24 & 104 \\
Jet veto & 41 & 43 & \multicolumn{1}{r}{$>557$} & 175 & 11 & 12 & 7.4 \\
$130 < M_T^{ll} < 160$ GeV & 18 & 11 & \multicolumn{1}{c}{---} & 0.21 & 1.3 &
0.04 & 0.09
\end{tabular}
\end{ruledtabular}
\end{table*}

The first level of cuts requires two isolated leptons, each with $p_{Tl}>10$
GeV and $|\eta_l|<2.5$.  Isolation of electrons and muons replicates recent
ATLAS descriptions \cite{ATLTDR,ATLLEPS}, and is applied within the modified
PGS detector simulation.  Next, a cut is placed on the transverse energy of
the reconstructed highest-$E_T$ lepton $l_1$ of $E_{Tl_1}>20$ GeV.  A fairly
high missing energy of 40 GeV in then required.  Since the leptons from $H\to
WW$ tend to go in the same direction, the invariant mass is low, and ATLAS
requires $M_{ll}< 80$ GeV.  The angle between the leptons should also be
small, and an aggressive cut is made on the azimuthal angle between them,
$\Delta\phi<1.0$.  Next, the dilepton system is required to be forward, with
the cut $|\theta_{ll}|<0.9$ (equivalent to $|\eta_{ll}|>0.73$).  Then, the
leptons are required to be close in pseudorapidity $|\eta_{l_1}-\eta_{l_2}| <
1.5$.  The next-to-last cut is a veto of any event with a jet with $E_{Tj}>15$
GeV, and $|\eta_j|<3.2$.  This cut serves to reject background from $t
\bar{t}$ production.  Finally, a tight cut is made on the transverse mass of
the dilepton and missing energy that naively appears to remove most of the
heavy-flavor background.

\begin{figure}[tbh]
\centering
\includegraphics[width=3.25in]{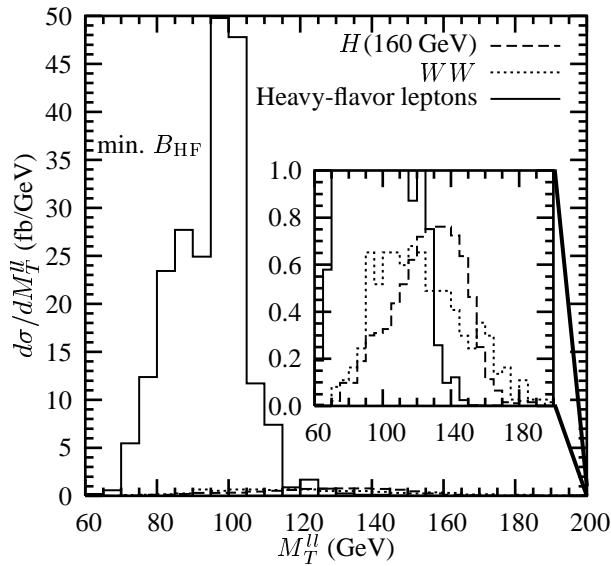}
\caption{Opposite-sign dilepton transverse mass for a 160 GeV Higgs boson, the
continuum $WW$ background, and the sum of additional heavy-flavor backgrounds
(HFB) at ATLAS.  The HFB is a lower limit on $b \bar{b}$ and does not include 
$c\bar c$.
\label{fig:atlmtll}}
\end{figure}

The most important observations regarding the ATLAS cuts are: the isolation is
roughly as effective at ATLAS as at the Tevatron experiments; and the final
cut on transverse mass $M_T^{ll}$ is the key to the ATLAS sensitivity.  In
Fig.\ \ref{fig:atlmtll} we see a comparison of the Higgs boson signal, the
continuum $WW$ background, and the new heavy-flavor backgrounds.  The new
backgrounds are more than an order of magnitude larger than the previously
calculated backgrounds for transverse masses less than 110 GeV.  This
observation is important, because the ATLAS transverse mass cut is rather
aggressive in the assumption that one can rely on the signal and background
distributions to claim discovery when the peak of the distribution is below
the cut.  Contrast this with \DZero, which uses $m_h/2$ as a lower limit.  A
\DZero-like cut would make $S/B\alt 1/30$ for a Higgs boson with mass less
than 200 GeV.

\begin{figure}[tbh]
\centering
\includegraphics[width=3.25in]{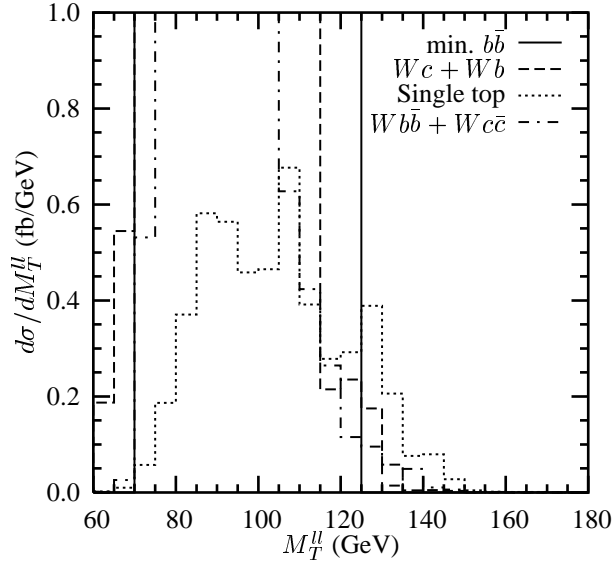}
\caption{Opposite-sign dilepton transverse mass for individual heavy-flavor
backgrounds (HFB) at ATLAS.  The $b\bar b$ background is a lower limit.
$c\bar c$ could be at least as large as $b\bar b$ production.
\label{fig:atlmtsplt}}
\end{figure}

\begin{figure}[tbh]
\centering
\includegraphics[width=3.25in]{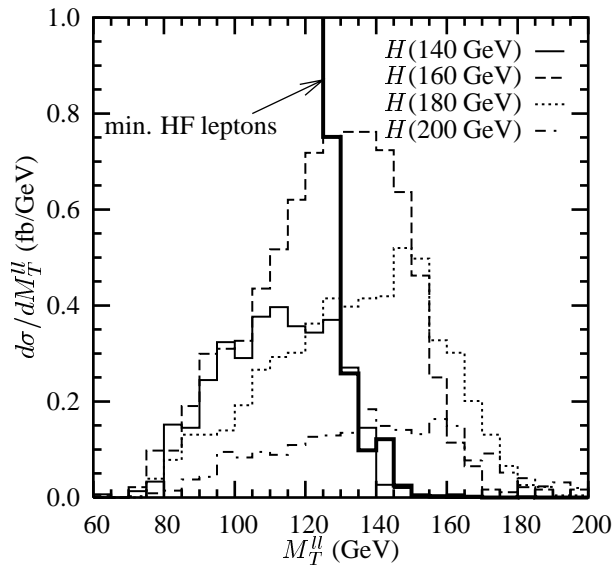}
\caption{Opposite-sign dilepton transverse mass for Higgs bosons at ATLAS.
\label{fig:hatatl}}
\end{figure}

The tail that creeps up to 150 GeV comes mostly from $t$-channel
single-top-quark production, which is fairly-well modeled.  In Fig.\
\ref{fig:atlmtsplt} we see the breakdown of the contribution of the HF
backgrounds.  As a result of the physics cuts and lepton isolation, the $b\bar
b$ background (and $c\bar c$ if it were included) has been suppressed by 11
orders of magnitude.  It is unlikely that the tail of that distribution cuts
off sharply at 125 GeV.  It would be difficult to believe any excess observed
in the region $M_T^{ll}<160$ GeV without a measurement of this HF background.
We see in Fig.\ \ref{fig:hatatl} that even a 200 GeV Higgs boson has a median
transverse mass below 140 GeV, leading to poor mass resolution even if events
are observed.

\begin{figure}[tbh]
\centering
\includegraphics[width=3.25in]{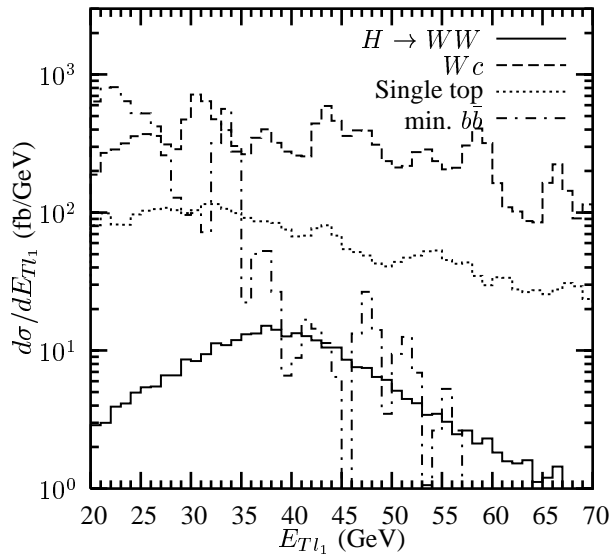}
\caption{Transverse energy distribution of the leading lepton after isolation
at ATLAS.
\label{fig:atletlo}}
\end{figure}

Fortunately, we can reduce the HF background to a manageable level by pushing
the $M_T^{ll}$ mass peak associated with the heavy-flavors below 110--120 GeV.
Applying lessons learned from our examination of the \DZero\ and CDF analyses,
we look again at the lepton transverse energies.  Figure \ref{fig:atletlo}
shows that the leading lepton in the $W+$jets and single-top samples is fairly
insensitive to small increases in the $E_T$ threshold near 20 GeV.  This is
not surprising, since the leptons come from real $W$ decay.  The leading
lepton from $b$ or $c$ decay falls faster, but an increase in the cut will not
improve the overall significance.  It may be desirable to raise the cut to
further suppress the poorly modeled $b\bar b$ and $c\bar c$ components, but
there is currently no clear gain from doing so.

\begin{figure}[tbh]
\centering
\includegraphics[width=3.25in]{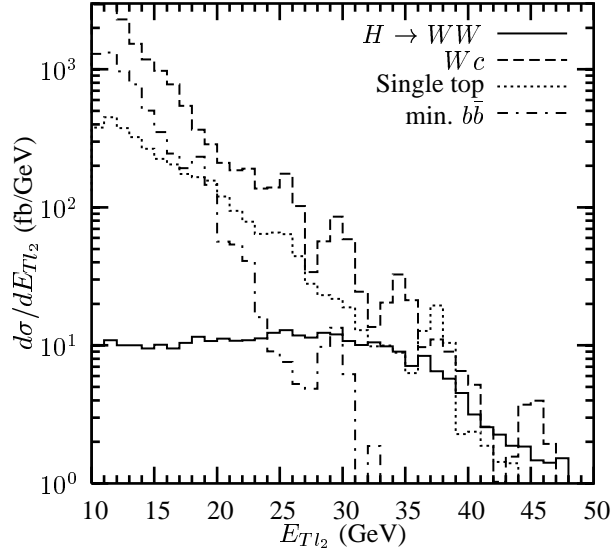}
\caption{Transverse energy distribution of the next-to-leading lepton after
isolation at ATLAS.
\label{fig:atletlt}}
\end{figure}

\begin{figure}[tbh]
\centering
\includegraphics[width=3.25in]{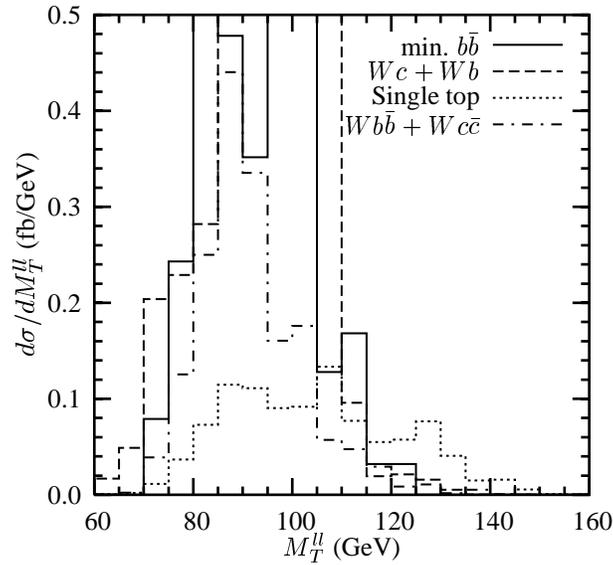}
\caption{Opposite-sign dilepton transverse mass for individual heavy-flavor
backgrounds (HFB) at ATLAS after the cut on $p_{Tl_2}$ is raised from 10 GeV to
20 GeV.  The $b\bar b$ background is a lower limit; the $c\bar c$ contribution 
could be as large as $b\bar b$ production.
\label{fig:atlmtllcutb}}
\end{figure}

\begin{figure}[tbh]
\centering
\includegraphics[width=3.25in]{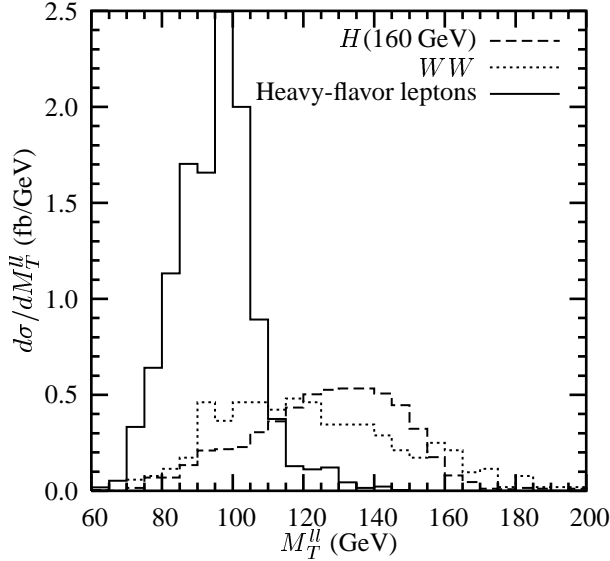}
\caption{Opposite-sign dilepton transverse mass for a 160 GeV Higgs boson, the
continuum $WW$ background, and the additional heavy-flavor background (HFB) at
ATLAS after the cut on $p_{Tl_2}$ is raised from 10 GeV to 20 GeV.
\label{fig:atlmtllcut}}
\end{figure}

The next-to-leading lepton in Fig.\ \ref{fig:atletlt} shows an exponentially
falling background as a function of $E_T$.  An increase of the minimum
transverse energy cut on additional leptons from 10 GeV to 20 GeV reduces the
background by roughly a factor of 20, while maintaining about $2/3$ of the
signal and continuum $WW$ backgrounds.  In particular, the dangerous $b\bar b$
background drops by a factor of 30, the $Wj+X$ backgrounds go down a factor of
10, and single-top-quark production goes down a factor of 5.  Such a cut is
nearly a ``magic bullet'' for Higgs boson masses above 140 GeV.  An estimate
of the effect of this one change in the cuts is shown for each of the
backgrounds in Fig.\ \ref{fig:atlmtllcutb}, and for the signal and total
backgrounds in Fig.\ \ref{fig:atlmtllcut}.  The leading edge of the
heavy-flavor transverse-mass peak is 20 GeV lower than with the default cuts.
The shift of this leading edge, along with the lower overall magnitude of the
background, protects the Higgs boson signal region from uncertainties in the
modeling of the heavy-flavor background.  The residual HF background will
still be measurable at lower $M_T^{ll}$, and it provides an \textit{in situ}
control sample.

While an increase in the lepton transverse-energy cut is effective at reducing
the background, it is important to note that every level of cuts is
significant.  In particular, some of the proposed cuts are potentially quite
sensitive to actual detector performance, noise, and the underlying event ---
none of which will be known until data are accumulated.  We point out the
effect on these backgrounds if some of the cuts have to change.

Since isolation criteria are not finalized, this seems a logical place to
start.  The complete prescriptions for isolation are described in Refs.\
\cite{ATLTDR,ATLLEPS}.  Electron isolation hinges on the energy $E_{12}$ being
less than 3 GeV in 12 calorimeter cells surrounding a central 2$\times$2 core.
Muons are required to have less than 4 GeV of additional charged tracks in a
cone of size $\Delta R=0.2$, and less than 10 GeV of calorimeter energy in a
cone of size $\Delta R=0.4$.\footnote{The tracking and calorimeter cuts for
muons are highly correlated.  Investigations into the efficacy of dropping one
cut are ongoing \protect\cite{ATLLEPS}.} The size of the underlying event and
measured shower shapes may require raising or lowering of thresholds for
allowed radiation.  Lowering the thresholds by a factor of two has little
effect on electron isolation, and only a factor of 2 reduction in the
background for muons.  Raising the isolation thresholds by 50\% (a typical
high-luminosity scenario~\cite{ATLTDR}) still has little effect on the
electrons, but increases the muon background by a factor of 2 per muon (so the
contributions from $b\bar b$ and $c\bar c$ increase by a factor of 4).

A change of the missing energy cut by $\pm 10$ GeV has little effect on the
overall significance.  However, the signal begins to decrease significantly
once the $\MET$ increases above 50 GeV.  The jet veto threshold is expected to
be raised to 40 GeV for high luminosity operation.  At a higher threshold, the
Higgs boson signal increases by about 25\%, while the background increases by
no more than 50\%.  It may be advisable to relax the jet veto threshold to 40
GeV to reduce sensitivity to the underlying event, and improve the
significance during low luminosity operation.  Finally, it is notable that the
tight $\Delta\phi$ cut reduces the signal by almost a factor of 2.  It may be
desirable to relax this cut to the \DZero\ choice $\Delta\phi<2.0$ to improve
the event rate.  None of these other cuts has nearly the impact of an increase
in the $E_T$ threshold.of the next-to-leading lepton.

\section{Conclusions}
\label{sec:conclusions}

In this paper we perform a full Monte Carlo simulation of the background for
Higgs boson decay to dilepton plus missing energy that arises from leptonic
decays of heavy flavors.  The processes that produce these backgrounds
typically begin $10^5$--$10^{12}$ times as large as the signals of interest.
Contrary to popular lore, these backgrounds are only mildly suppressed by
isolation cuts.  Instead, it is the detailed sequence of cuts on the phase
space of the events that suppresses their size.

Throughout this paper we compare the heavy-flavor background (HFB) to the
largest previously calculated background, continuum $W^+W^-$ production.  For
events with one heavy-flavor lepton ($Wc$, $Wb\bar b$, $\ldots$), the analysis
cuts tend to chip away at the heavy-flavor backgrounds.  The net result is an
additional background that is half as large as continuum $WW$ at the Tevatron,
but potentially $25$ times larger than $WW$ at the LHC.  For events with two
heavy-flavor leptons ($b\bar b$, $c\bar c$), the additional background
estimate is less certain, but it is potentially even larger.

Spin correlations in the $H\to W^+W^-$ signal tend to cause the leptons to be
fairly soft, hence the $E_T$ cut on the next-to-leading lepton is pushed as
low as possible --- 10 GeV in the cases of both \DZero\ and ATLAS.  We
demonstrate that simply raising the cut on this additional lepton to $E_T>$20
GeV preserves roughly $2/3$ of the signal, but it reduces these background by
factors of 10--30.

Raising the $E_T$ cut is very effective for removing the HFB, but we emphasize
that extrapolations of the magnitude and shape of this background using Monte
Carlo techniques have large inherent uncertainties.  In particular, there is a
strong correlation between the cut on missing transverse energy $\MET$ and
isolation.  We recommend that the HF background be measured \textit{in situ},
with cuts as close as possible to the final sample.  We describe a preliminary
technique that could be used at the Tevatron.  At the LHC the HFB is large
enough that it can be studied in the final transverse mass $M_T^{ll}$
distribution and fully controlled.

While this paper focuses in detail on analyses of $H\to W^+W^-$, our intention
is to raise a broader awareness of the potential danger of heavy-flavor
leptons to multilepton analyses.  Vector-boson fusion into a Higgs boson is
similar to the process studied here, with the added requirement of two hard
jets at large and opposite rapidities.  The jet requirement may
\textit{increase} the backgrounds from $b\bar b$ and $c\bar c$, because these
processes naturally come with additional radiation.  Other Higgs boson decay
modes may also be effected, e.g., $H\to ZZ\to 4$ leptons has a background from
$Zb\bar b$.  Even trilepton searches for supersymmetry typically have very
soft additional leptons.  If lepton transverse momentum cuts must be raised to
remove the heavy-flavor leptons, there could be a far-ranging impact on
analyses of these types of signals.  Complete correlated studies should be
performed to determine whether heavy flavor leptons are a problem and
contingencies made to measure them.

\begin{acknowledgments}
Z.\ S.\ is a visitor at Argonne National Laboratory and is supported in part
from an Argonne-University of Chicago Joint Seed Grant.  E.\ B.\ is supported
by the U.~S.\ Department of Energy, Division of High Energy Physics
under Contract No.\ W-31-109-ENG-38.
The authors thank Tom LeCompte, Jim Proudfoot, and Frank Paige for discussions
regarding the ATLAS detector, and John Campbell for help with MCFM.

\textit{Note added}--- As this paper was being completed, the CDF
Collaboration released a preprint \cite{unknown:2006aj} describing a new
analysis based on the \DZero\ method studied here.  In that paper, the $b\bar
b/c\bar c$ resonances are suppressed by the cut on the minimum dilepton
invariant mass $M_{ll}>16$ GeV.  According to our investigation, the minimum
$E_T$ cuts and weighting of the $\Delta\phi$ distribution force $M_{ll}>16$
GeV for most heavy flavor events (all events with $\Delta\phi>1.2$).  This
dilepton mass cut is looser than the \DZero\ choice of 20 GeV and will yield
less suppression against $Wc$.  The $W+$heavy-flavor events and
$b\bar b/c\bar c$ events may well make up the systematic difference between
the background estimate and the small excess observed in the CDF data.
\end{acknowledgments}

\end{document}